\documentclass[nonacm=true]{acmart}
\setcopyright{none}
\renewcommand\footnotetextcopyrightpermission[1]{} 
\makeatletter
\let\@authorsaddresses\@empty
\makeatother
\usepackage{hyperref}
\usepackage{enumitem}

\AtBeginDocument{%
  \providecommand\BibTeX{{%
    \normalfont B\kern-0.5em{\scshape i\kern-0.25em b}\kern-0.8em\TeX}}}

\begin{document}

\title[Sequential/Spatial]{Sequential/Spatial, a Survey of Interactive Information Retrieval Methods for Controlled Experimentation and Evaluation}

\author{Michael Segundo Ortiz}
\affiliation{%
 \institution{Carolina Health Informatics Program, University of North Carolina at Chapel Hill}
 \city{Chapel Hill}
 \state{NC}
 \country{USA}}
\email{msortiz@unc.edu}

\renewcommand{\shortauthors}{Ortiz}

\begin{abstract}
This survey presents studies that investigated non-spatial (sequential) and spatial information retrieval systems in parallel during a battery of information-seeking tasks with respect to user navigational behaviors, incidental learning, retrieval performance, cognitive abilities \& load, direct manipulation of 2D \& 3D interfaces, and satisfaction. I consider how information theory has contributed to the concepts of foraging, sense-making, exploration, and how the applied areas of interactive information retrieval (IIR) and cognitive/behavioral psychology have implemented these concepts into architecture, interface design, experimental design, user study, and evaluation methodology.
\end{abstract}

\maketitle
\thispagestyle{empty}

\section{Introduction}
Information retrieval engines work in a vast information space and time. These spaces grow exponentially while time moves linearly. The rate of output is not amenable to solely ranked-retrieval or sequential encoding. Moreover, this paradigm does not consider individual differences in cognitive ability or learning styles for accessing, retrieving, and comprehending information. So I ask this question: \textit{If an information space and our cognitive space are very separated (e.g. I would like to learn more about genomic editing technologies and the corresponding research) how do current information retrieval systems enable us to explore the space?} 

As stated by Benyon and H{\"o}{\"o}k \cite{Benyon1997NavigationII}, \textit{How do people work out how to reach their destination? The answers to this question are many and, often, unsurprising. People use maps and guides..landmarks}. Thus, in many ways the notion of \textit{Sequential/Spatial} or taking sequential information and encoding it in a spatial context can be regarded as information cartography \cite{shahaf2013information}. On this matter, I am not concerned with the visualization component per se. Rather, I am focused on how the human is measured while navigating information that is enabled by the visualization, both from retrieval performance and cognitive/behavioral, perspectives. Moreover, all spatial-semantic information retrieval system architectures are \textit{visualizations} however, not all \textit{visualizations} are spatially and semantically encoded. Thus, for experiments that have humbly motivated this survey, albeit not included due to highly focused criteria, please examine \cite{pirolli1996scatter,westerman2000cognitive,westerman2000mapping,westerman2005browsing,ruotsalo2013directing,gonzalez2017effects,peltonen2017topic,schleussinger2018evaluating,zhang2014evaluation,koshman2004comparing,klouche2017visual,ruotsalo2018interactive}.

There are of course various types of metaphors that can be used for the architectural design and navigation of an information space. In terms of compartmentalizing results displayed by an information system or digital library more specifically, multifaceted search is one option. Take for example biomedical literature. As currently implemented by the National Center for Biotechnology Information (NCBI)\footnote{\url{https://www.ncbi.nlm.nih.gov}}, the architecture represents biomedical information objects in terms of literature, genetics, proteins, and chemicals. More granular facets exist within each information object. One can apply facets for article type, text availability, publication date, species studied, ages studied, etc. However, even with the application of facets, search results are still sequentially encoded according to relevance or date by default, and can range from the thousands to millions of relevant results without any metaphorical presentation of the concept space \cite{zhang2002information}. This elicits a question. \textit{How does a user establish a mental model of semantic structure within a space that is extraordinarily large and must be explored with nothing more than a list with filters? Of equal importance, how are experiments designed in order to answer this question?}

It is an important investigation as to what alternative modalities of information access exist and/or are possible that enable an information seeker to locate objects, generalize, infer, cluster, analyze, and learn efficiently, which forms the basis of a \textit{LOGICAL} human-computer information retrieval framework that may reduce the theoretical distance between an information space and our own cognitive space. In other words, the continuous rate of change and a currently limiting mode of access, may lead to wider gaps in the pipeline of information discovery and comprehension.

For system-level architectures, their currently exists two modalities of text retrieval: sequential (non-spatial) and spatial. In this review I omit voice-based information retrieval for smart-phone devices and restrict ourselves to the personal computing platform, whether on a laptop or desktop. For this review, I have only one general research question (RQ):

\begin{itemize}
  \item[] \textbf{RQ1:} What previous work exists that systematically investigates the differences between sequential and spatial information retrieval architectures and their effectiveness for an information seeking task?
\end{itemize}

The following criteria (C) must have been met to incorporate a publication for review:

\begin{itemize}
  \item[] \textbf{C1:} Must empirically examine the effectiveness of sequential and spatial information retrieval architectures in parallel.
  \item[] \textbf{C2:} The evaluation must be conducted in an experimentally controlled manner using control and experimental groups of human users.
  \item[] \textbf{C3:} Elements of cognitive and behavioral psychology must be measured.
  \item[] \textbf{C4:} Information seeking must be the primary task.
\end{itemize}

The rest of this review is organized as follows. In section 2, I briefly introduce the concept of \textit{Sequential/Spatial} from a historical context. In section 3, I dive deep into the experimental designs and findings. In section 4, I discuss common themes, problems, insights, and how they fit together to inform the reader on what gaps might exist. In section 5, I provide my concluding remarks.

\section{Sequential/Spatial}
In the early days of information retrieval system interfaces, like those produced by Digital Equipment Corporation, later acquired by Compaq, designs were very simple. It was argued that as the complexity of stored information grew, human mental models of where information was located, and how to arrive at the information target, would degrade. Moreover, pictorial presentations of data would aid in the initial comprehension of complex data or information spaces, and provide a visual reference point that could be returned to repeatedly. Thus, architectures that enabled navigation of computer menus via a supplemental visual map were explored \cite{billingsley1982navigation}. The architecture I refer to here was quite literally a map printed on legal size paper, held horizontally, and depicted a tree diagram of menu options. In the early 1980's graphical or pictorial representations of data structures for information retrieval systems were growing in interest. However, there was a lack of empirical evidence that existed to support or refute the use of visual aids.

New questions were asked in the mid 1990's as personal computers, and World Wide Web (WWW) use, more than doubled \footnote{\href{https://www.bls.gov/opub/btn/archive/computer-ownership-up-sharply-in-the-1990s.pdf}{https://www.bls.gov/opub/btn/archive/computer-ownership}}. It was likely that people would now encounter unknown digital information spaces. Moreover, the formulation of a query was non-trivial and often vague. Thus, the argument of how to develop an architecture that communicated topic structure of text collections arose \cite{pirolli1996scatter}. Xerox Parc led this effort. The paradigm was called \textit{Scatter/Gather}. This was an important development because an architecture that allowed computerized iterative navigation of topics and their clustered texts afforded users the possibility of exploring concept spaces and learning what was inside before specifying a precise information target. The digital table-of-contents metaphor was born.

As the late 1990's approached and graphical user interfaces (GUIs) became more sophisticated along with Graphical Processing Units (GPUs), 3-dimensional visualizations of information spaces that represented document relatedness grew in architectural popularity. This was primarily due to the push by the National Institute of Standards and Technology Text Retrieval Conference, \textit{NIST-TREC}\footnote{\url{https://trec.nist.gov}} for short, and their emphasis on \textit{Aspect-Oriented Information Retrieval}. Similar to \textit{Scatter/Gather}, the \textit{2D and 3D Aspect Window} architectures attempted to convey textual topology albeit with added dimensionality, which as I will investigate later, added cognitive demand \cite{swan1997aspect,sebrechts1999visualization,koshman2004comparing}.

At this point in the time-line of information retrieval systems, no experimental architecture had explored the relationship between topology and semantics. The advent of spatial and semantic navigation was vital to the information retrieval field. A simple question arose several years after automatic indexing by Latent Semantic Analysis (LSA) was first introduced \cite{deerwester1990indexing}. \textit{Does spatial distance equal semantic distance?} Chen proposed a system architecture that addressed this question \cite{chen2000individual}. It became clear that such an architecture could help users understand where information objects were, what they meant, what they were related to, and more importantly to what degree. 

Between the years 2004 and 2016 there was a relative academic silence with respect to architectural prototypes and the comparative evaluation of \textit{Sequential/Spatial}. Some important works on \textit{Exploratory Search} were published although most of them were position pieces or non-comparative with respect to \textit{"sequential versus spatial"} in their experimental methods. Some of the more impactful publications on this matter can be found here \cite{marchionini2006exploratory,capra2008relation,white2008evaluating,white2009exploratory,kules2009exploratory,ruotsalo2013directing}. Interestingly, although merely correlative, these years also aligned with two important events for the information science world that included Google's supremacy in the information services market as they announced their initial public offering (IPO) in 2004, followed by the United States presidential election of 2016, the rise in misinformation, and a highly-charged unstable geopolitical landscape. I mention these sociocultural events not to take a political position but rather to attempt reason for the lack of experimental and comparative evidence in the area of \textit{Sequential/Spatial} evaluation followed by the subsequent rise in 2017. 

In 2017 sequentially encoded information retrieval systems were still the state-of-the-art. One-dimensional ranked lists were provided as search engine result pages (SERPs). A seemingly contradictory approach, as a gold standard, seemed absurd to many.\textit{What sense did it make to provide one-dimensional results for multi-dimensional queries? Additionally, how does a user decide which information objects are relevant to which query terms?} The problem was far beyond a vague and ambiguous query or an untraversed information space needing to be visualized for the purposes of having a map, table-of-contents, or spatial-semantic, metaphor. It was now the problem of information seeking intent, interactive feedback, iterative browsing, and search-as-learning. The notion of \textit{Multi-Aspect Information Retrieval} revived previous work although under a slightly different approach and name \cite{klouche2017visual,peltonen2017topic,schleussinger2018evaluating, ruotsalo2018interactive}.

\section{Experimental Designs and Findings}
Patricia Billingsley \cite{billingsley1982navigation} designed an experiment that had three research questions. They are paraphrased below.

\begin{itemize}
  \item By studying a linear index of correct information system menu options, would a user improve their speed and accuracy on an information seeking task?
  \item Are spatial-semantic maps of information system menu options more effective for information seeking tasks than lists that provide only the semantic associations?
  \item Is there a relationship between individual cognitive ability and information retrieval performance?
\end{itemize}

Male and female undergraduate/graduate psychology students were the participants of the study and were given a cognitive screen provided by Educational Testing Service (ETS)\footnote{\url{https://www.ets.org}} that assessed spatial and verbal memory. A database was created specifically for the experiment and included animals with their descriptions that varied in difficulty and ambiguity. Prior to the experiment instructions were read aloud to the participants followed by an example task. The experiment had five trial blocks. The first three trial blocks were considered a learning phase and therefore had the same set of target animals whereas trial blocks four and five had a different set of target animals. After trial block one, the experimental groups were allowed to study their index or map for a few minutes while the control group continued with the task. The control group studied no index or map prior to the information seeking task.

\begin{figure}[htb]
    \centering
    \includegraphics[width=7cm]{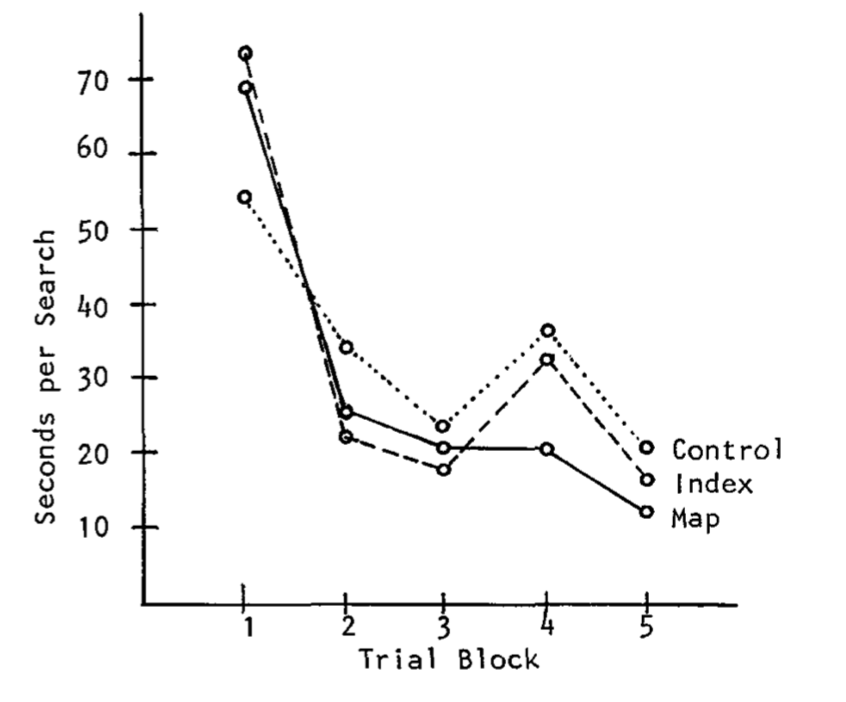}
    \caption{Average time.}
    \label{fig:fig1}
\end{figure}

\begin{figure}[htb]
    \centering
    \includegraphics[width=7cm]{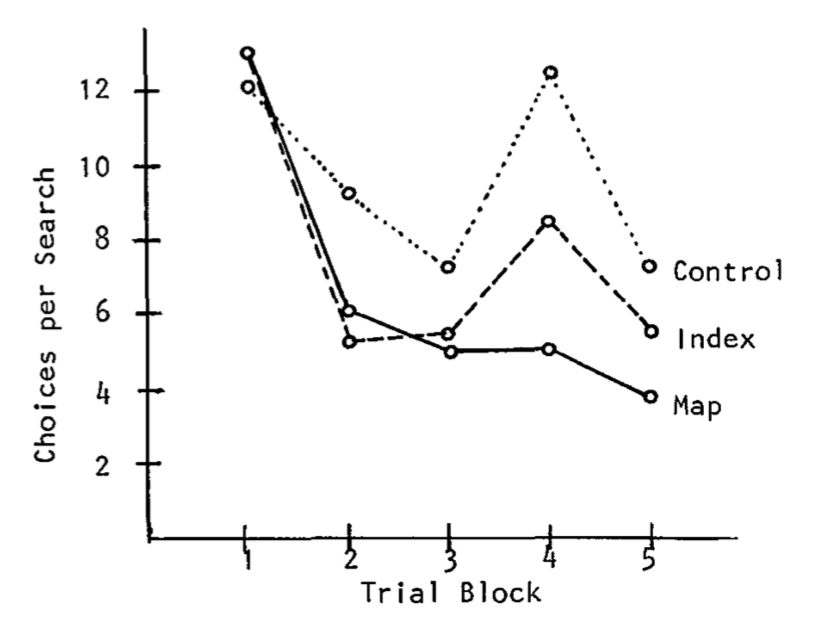}
    \caption{Average number of choices.}
    \label{fig:fig2}
\end{figure}

Performance was examined by keeping track of the time and specific sequence of choices users made in comparison to the minimum number of required choices to reach the information target. It was observed that the spatial-semantic map-based experimental group not only reached the information targets in half the time but also made half the number of choices to reach the target. Figure \ref{fig:fig1} displays results for average time for each search by experimental and control groups across trial blocks. Figure \ref{fig:fig2} displays results for average number of choices per search by experimental and control groups across trial blocks. Both Figures have been directly extracted from the original publication.

Not only did the spatial-semantic map-based experimental group perform significantly faster (p<.01) in comparison to the control group, this group also made significantly fewer choices to target (p<.02). It was also observed that spatial memory from ETS was correlated with both the total time and number of choices.

Swan and Allan \cite{swan1997aspect} were interested in building and evaluating information system architectures that enabled \textit{aspect-oriented information retrieval} as outlined by the NIST-TREC-6 Interactive Track\footnote{\url{https://trec.nist.gov/data/t6i/t6i.html}}. The essential goal of the TREC Interactive Track was to develop methodology that not only evaluated performance outcomes of information-seeking tasks but also the process by which human cognition achieves the outcome. To examine information processing, the authors developed four research questions that are paraphrased below.

\begin{itemize}
  \item Can information systems with enhanced features such as \textit{aspect} keywords or spatial encoding be built and outperform a sequential system?
  \item Can indirect comparisons be made by implementing an effective control system?
  \item What are the cognitive and demographic factors that distinguish effective information system use for different architectures?
  \item Is 3-D visual-spatial encoding useful for information-seeking?
\end{itemize}

Three information systems were used for the experiment. ZPRISE (ZP)\footnote{\url{https://www-nlpir.nist.gov/works/papers/zp2/zp2.html}} was developed by NIST as a publicly available and sequentially encoded IR system which was used as the control system, AspInquery (AI) was developed by the authors and also sequentially encoded but with an added \textit{aspect} window that includes keywords to characterize document relevance, and AspInquery Plus (AI+) which was an extension of AI but with an \textit{aspect} window that encoded the information space in 3-D to present document relatedness.

\begin{figure}[htb]
    \centering
    \includegraphics[width=7cm]{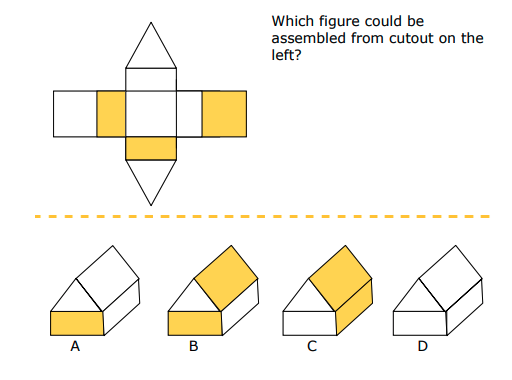}
    \caption{Example of a spatial reasoning test. The correct answer is B.}
    \label{fig:fig5}
\end{figure}

There were a total of 24 participants in the study. Half were librarians with Masters of Library Science degrees (MLS), while the other half represented undergraduate/graduate students with various backgrounds. Each experimental block had four users, and each user queried six topics. Three topics were queried on the control system (ZP), and three were queried on the experimental systems (AI and AI+). All participants filled out a demographic questionnaire and completed psychometric testing for verbal and spatial ability prior to experimentation. Verbal ability was assessed by the \textit{Controlled Associations} test which seeks to characterize how a subject establishes connections of relevant ideas that result from a specific stimulus. Spatial ability was assessed by the \textit{Paper Folding} test which attempts to characterize how a subject interprets 2-D drawings from mental images and visualizes their movement or changes. An example of a spatial reasoning test\footnote{\url{https://practicereasoningtests.com/mechanical-reasoning/}} is shown in Figure \ref{fig:fig5}.

The experiment was conducted in single-blind format where the participants were not told which system (control or experimental) they were exposed to until the post-experiment debrief. The dataset encompassed 200,000 newspaper articles from the Financial Times between the years 1991-1994 which represented a subset of the TREC collection. The search task was timed in all conditions with the time limit being 20 minutes. Each search task had an \textit{aspect} as defined by NIST assessors. For user performance, their saved documents during the search task were assessed by experts and then time-to-search, aspectual precision, and recall scores were measured. 

\begin{figure}[htb]
    \centering
    \includegraphics[width=7cm]{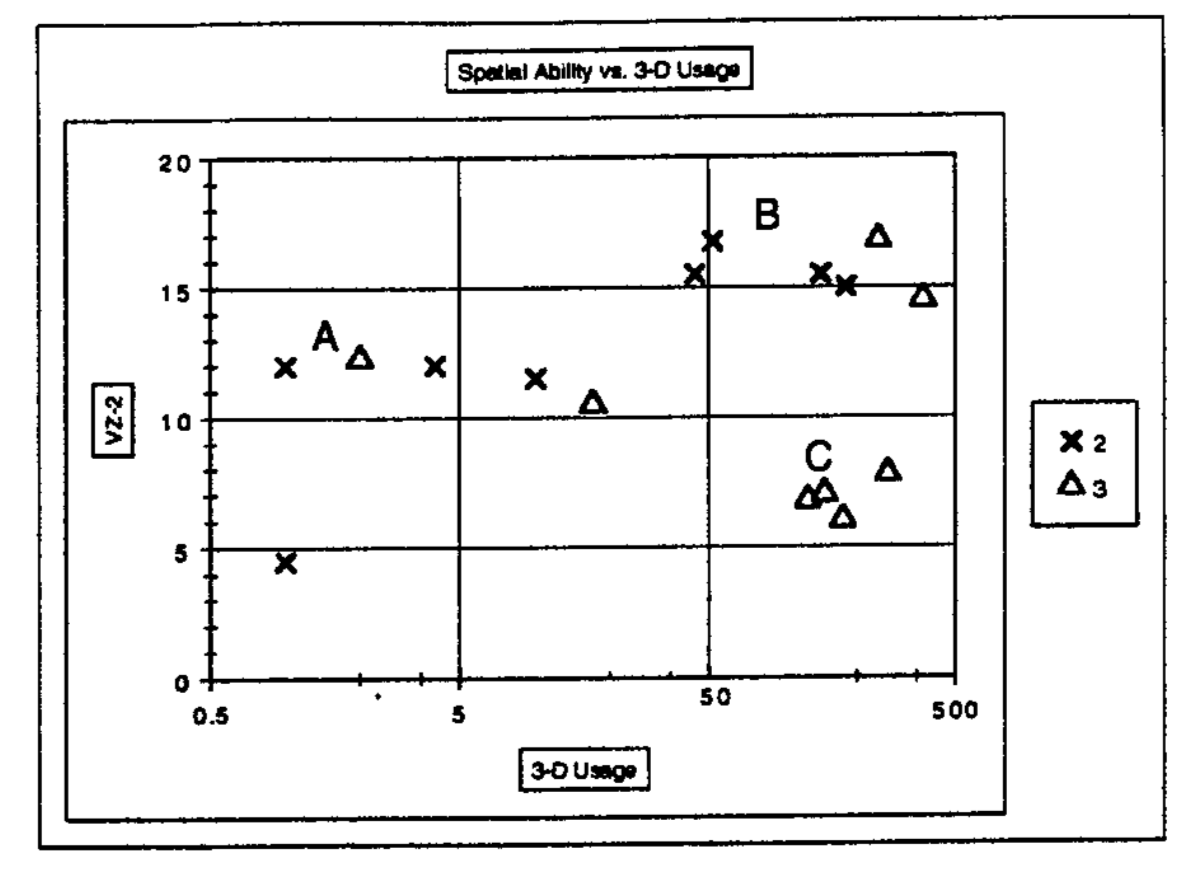}
    \caption{Cluster A: low 3D \textit{aspect} window usage BUT high spatial reasoning scores (VZ-2), Cluster B: high 3D \textit{aspect} window usage AND high spatial reasoning scores (VZ-2), Cluster C: high 3D \textit{aspect} window usage BUT low spatial reasoning scores (VZ-2)}
    \label{fig:fig6}
\end{figure}

In the experiments, user spatial ability did not appear to contribute significantly, although the sample size was small. Moreover, they observed that the best predictor of what a user will engage with (sequential or spatial) while actively using the experimental systems, is prior experience with similar elements in the past. This would suggest that a learning phase could be implemented before experimentation. On the issue of direct manipulation of 3D virtual environments with a 2D input device (computer mouse), the authors wanted to understand the relationship between spatial reasoning psychometric performance and the ability to engage in direct manipulation of a virtual environment. It was observed that although some participants achieved high spatial reasoning scores, they did not all engage with the AI+ 3D \textit{aspect} window as shown in Figure \ref{fig:fig6}. The authors administered a questionnaire upon these findings and concluded that familiarity with Graphical User Interfaces (GUIs) is more indicative of effective use than spatial ability.

To determine the effectiveness of the experimental systems (AI and AI+) against the control (ZP), the authors performed an analysis of variance with all interactions. Search topic, searcher, and system architecture were treated as independent variables while \textit{aspect} precision, recall, and time, were treated as dependent variables. The spatially encoded experimental system (AI+) outperformed the control system (p < 0.06) in recall. Although the control system outperformed the \textit{aspect-oriented} and non-spatial experimental system (AI) in recall (p < 0.04). As a recall enhancing device by design, the 3D spatial encoding was observed to be measurably useful. All other dependent measures were found to be not significant across conditions. From the independent variables, topic was the most significant predictor of recall, precision, and time-to-search. It is well known that topic difficulty plays a significant role in IR tasks, thus this result is not surprising. The least predictive factor was the system architecture or interface design.

\begin{figure}[htb]
    \centering
    \includegraphics[width=7cm]{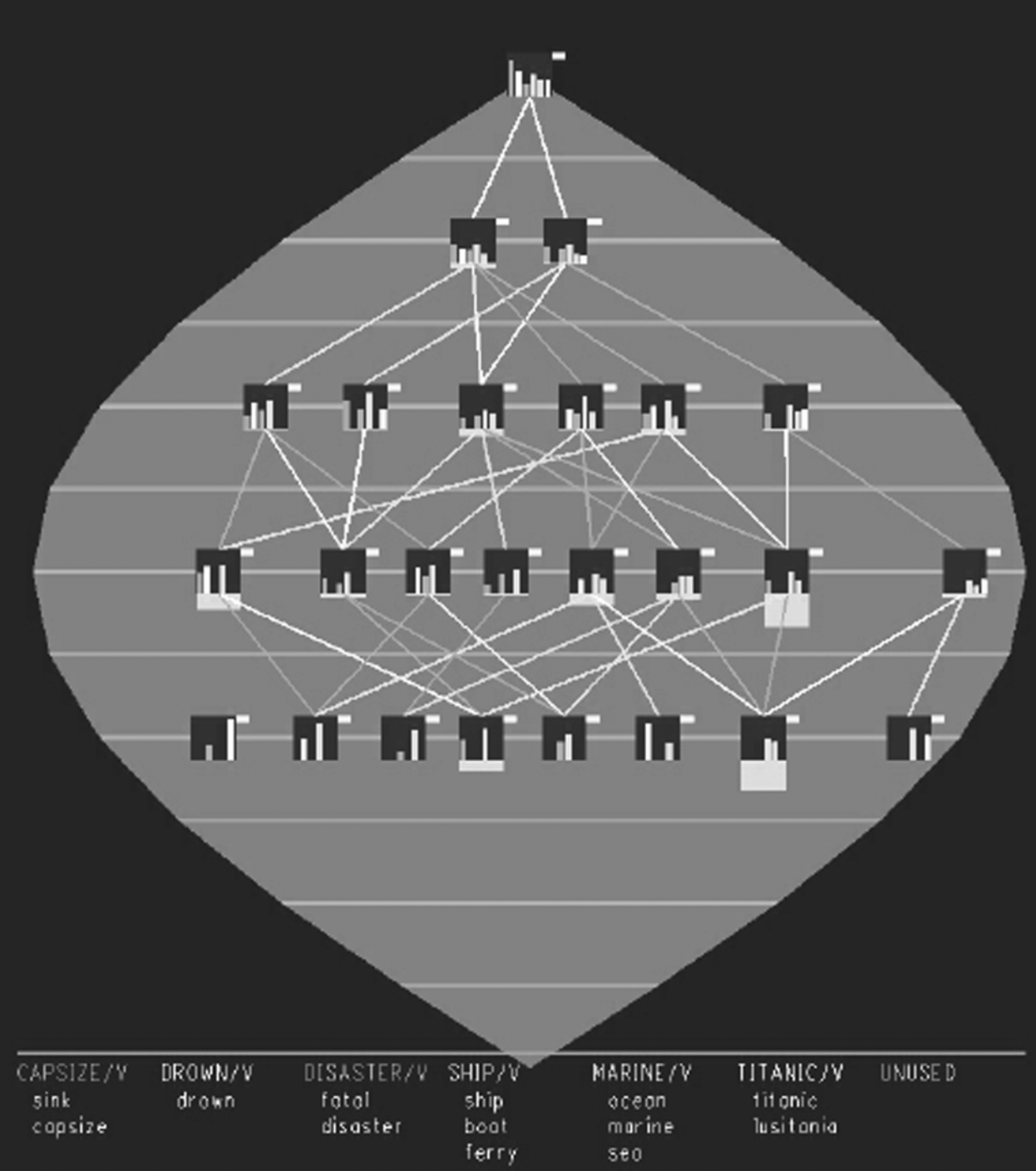}
    \caption{2D spatial encoding of the NIRVE information space. Top of the hierarchy represents the most relevant document cluster. Straight lines (edges) encode a degree of relatedness. Histograms encode keyword frequencies.}
    \label{fig:fig7}
\end{figure}

\begin{figure}[htb]
    \centering
    \includegraphics[width=7cm]{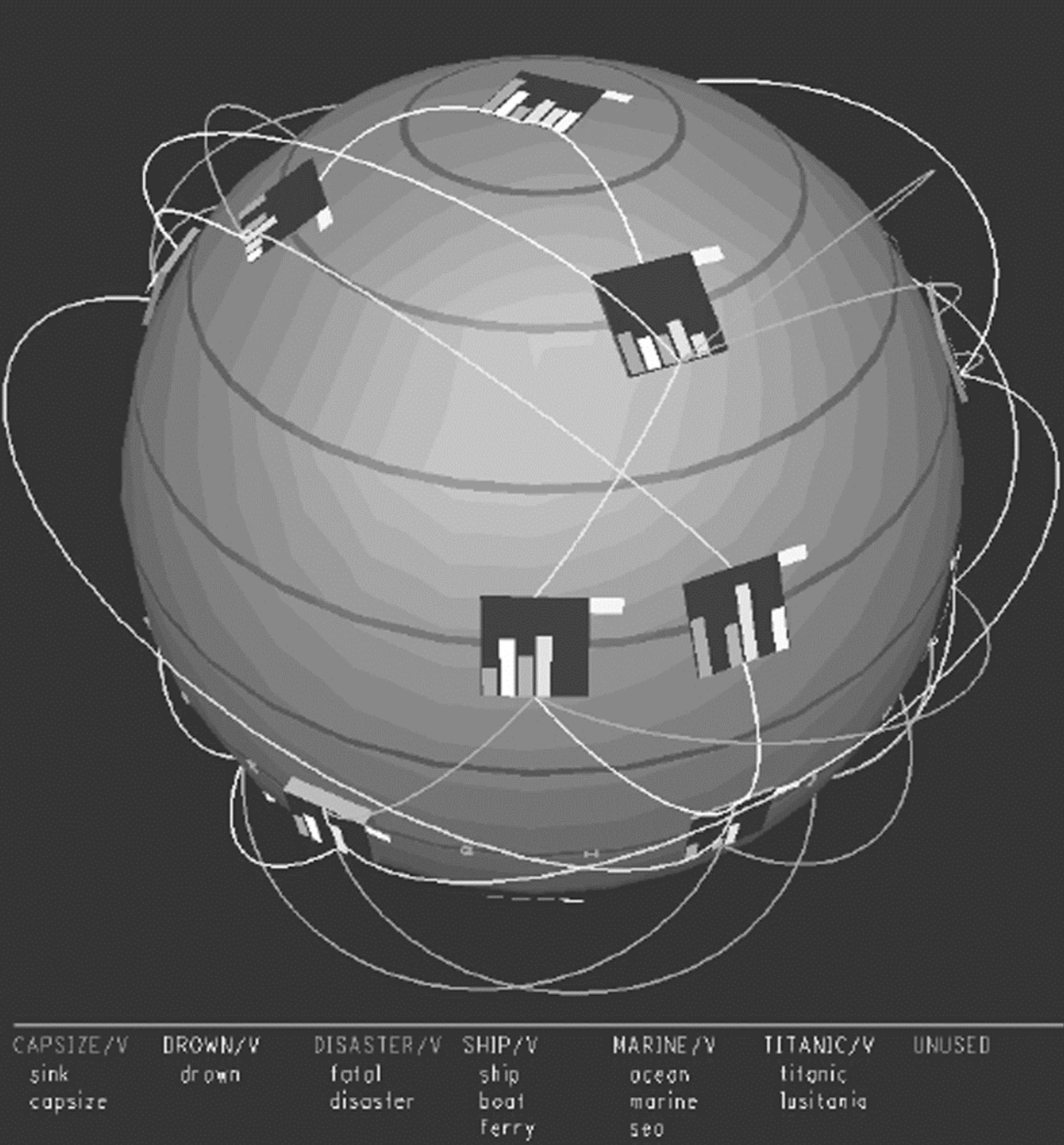}
    \caption{3D spatial encoding of the NIRVE information space. Most relevant documents are located at the 'North Pole' of the spherical layout. Arcs (edges) encode a degree of relatedness. Histograms encode keyword frequencies.}
    \label{fig:fig8}
\end{figure}

\begin{figure}[htb]
    \centering
    \includegraphics[width=7cm]{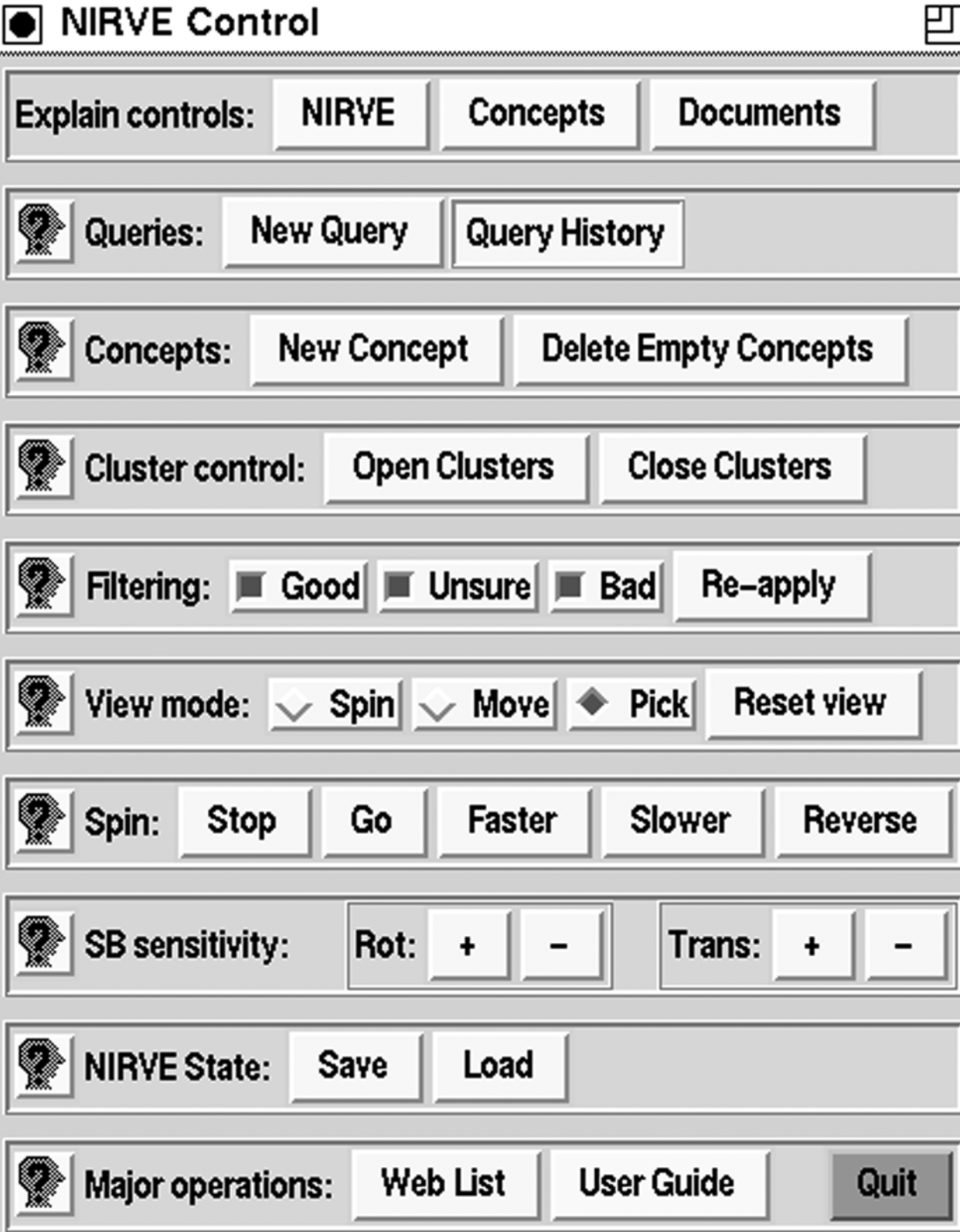}
    \caption{Control panel for 2D and 3D NIRVE architectures.}
    \label{fig:fig9}
\end{figure}

Sebrechts et al. \cite{sebrechts1999visualization} considered the notion of complicated information spaces and perhaps how spatial context could be used to decrease task demand during information seeking. The authors also considered the paradox of increasing information visualization practices but the lack of systematic comparison on the value of such practices. The study attempted to provide a controlled comparison of text-based, 2D, and 3D retrieval architectures for a task involving a small collection of 100 documents on the NIST Information Retrieval Visualization Engine prototype (NIRVE). The experiment was slightly different from others mentioned in this review in that users did not query but rather were given predefined queries based on previous TREC topics for a subset of 1988 news articles from the Associated Press. The spatial encoding of the 2D and 3D experimental systems can be seen in Figures \ref{fig:fig7} and \ref{fig:fig8}. For both spatially encoded systems, a control panel (Figure \ref{fig:fig9}) was implemented that allowed various functions such as filtering, relevance judgements, etc. 

The control system for the experiment was the sequentially encoded PRISE system developed by NIST, just as we previously discussed in the Swan and Allan experiments. The fifteen participants of this study were categorized as novice or professional. All were university students with/without significant computer experience. All participants completed questionnaires that provided demographic information and self-reported computer skills. In order to make an attempt at generalization, there were a variety of topics and tasks in the experiment. For each topic, participants had to conduct sixteen tasks that included locating, comparing, and describing documents or clusters of documents. A few examples of these tasks are paraphrased below.

\begin{itemize}
  \item Locate and skim an exact article and mark it in the control panel so you can find it later.
  \item Retrieve the marked article and locate an article with a similar title.
  \item Locate all clusters that discuss keyword 1 and keyword 2.
\end{itemize}

\begin{figure}[htb]
    \centering
    \includegraphics[width=7cm]{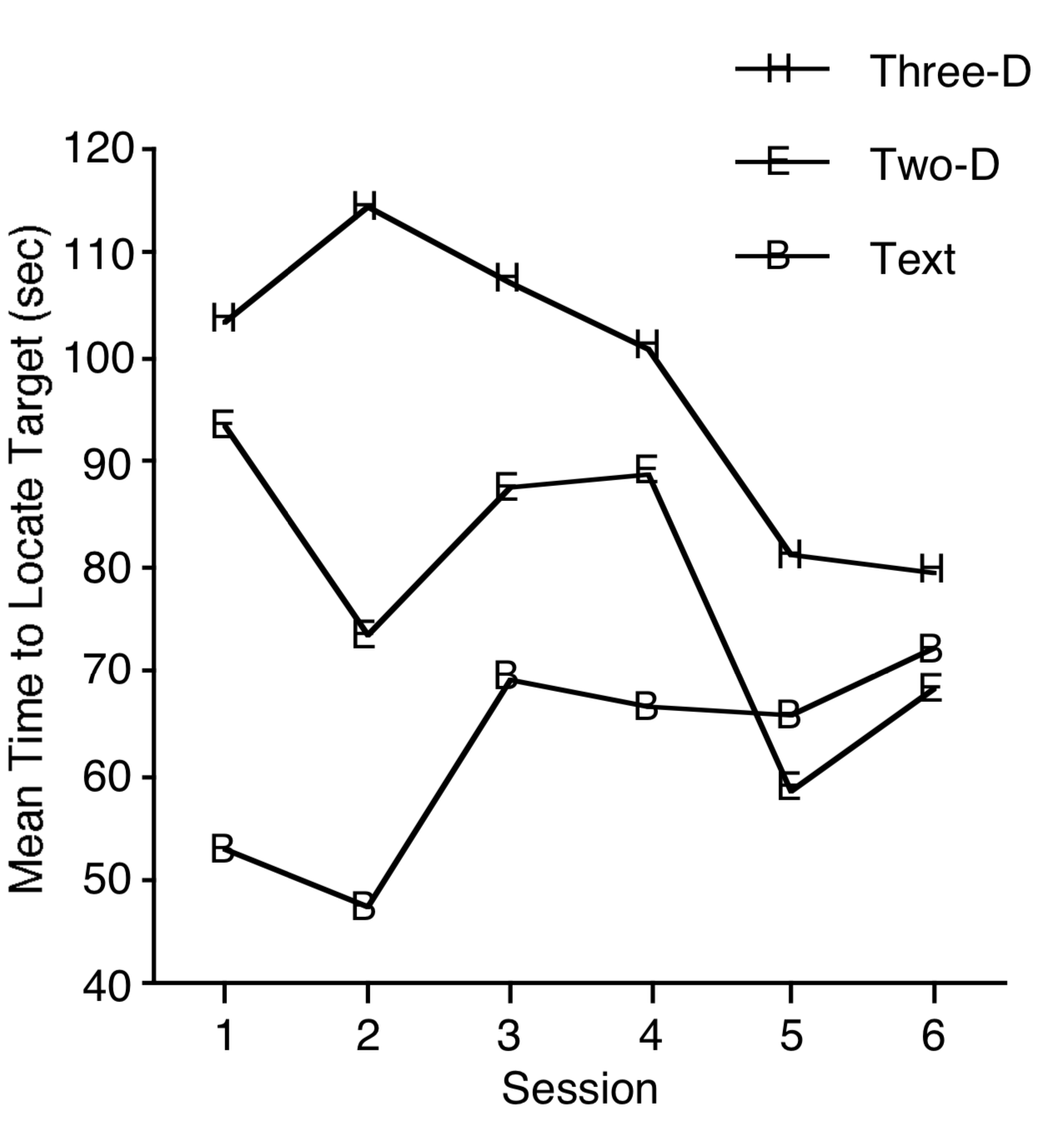}
    \caption{Average time-to-target across sessions and by system architecture.}
    \label{fig:fig10}
\end{figure}

Participants were given a brief tutorial on system interaction techniques by the experimenter. During the experiment, which lasted approximately one hour over three consecutive days, participants had a time limit of three minutes for each information seeking task and were asked to 'think-aloud' their thought processes for the duration of the task while the audio was recorded. A hint, and one additional minute was given if the participants did not finish at the time limit. After each topic and battery of tasks, a five minute break was given. Overall, the text-based control group performed their tasks faster across all topics (sessions) except sessions 5 and 6 where the 2-D architecture enabled slightly faster retrieval, as is shown in Figure \ref{fig:fig10}. However, as the experimental sessions continued, the time-to-target appeared to converge across all conditions. Once again, as we have described in the previous experiments, the learning curve for spatial metaphors appears to be rather quick, although in this experiment no psychometric evaluation was administered. Therefore, I am cautious in my assessment of what role cognitive spatial ability may have played in this learning curve. 

\begin{figure}[htb]
    \centering
    \includegraphics[width=7cm]{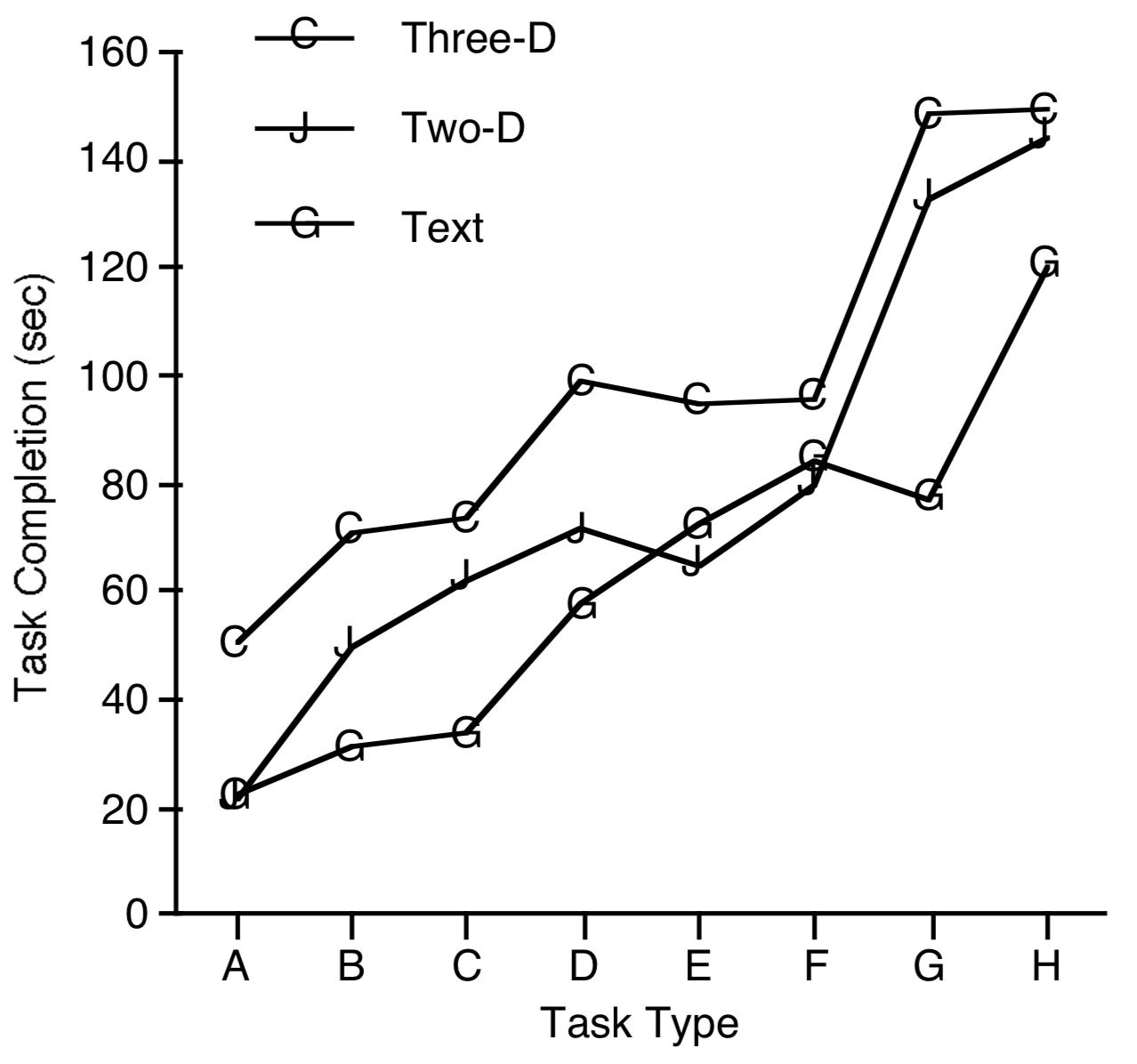}
    \caption{Average time-to-target by system architecture and task type.}
    \label{fig:fig11}
\end{figure}

There was a reliable difference in response time across the different task types as shown in Figure \ref{fig:fig11}. The task types are itemized below and have been directly taken from the original publication.

\begin{enumerate}[label=\Alph*]
\item Locate a cluster given its concepts.
\item Locate a document given its concepts \& title.
\item Locate a document given its concepts.
\item Recover a document given its title and follow a link to a new document.
\item Recover a document and locate a new document given its title or content.
\item Recover and compare contents of documents.
\item Determine the concepts for a document and locate it given its title.
\item Determine the concepts for a document and locate it given its contents.
\end{enumerate}

Overall, it was observed that determining a documents key concepts and then locating it within an information space required a significant investment in time for all system architectures (text,2D,3D). Moreover, computer experience also played a critical role for 3D retrieval performance but not for text or 2D based architectures, suggesting that 3D presentation demands higher cognitive effort due to direct manipulation of a 3D space with a computer mouse that functions on a 2D plane.

\begin{figure}[htb]
    \centering
    \includegraphics[width=7cm]{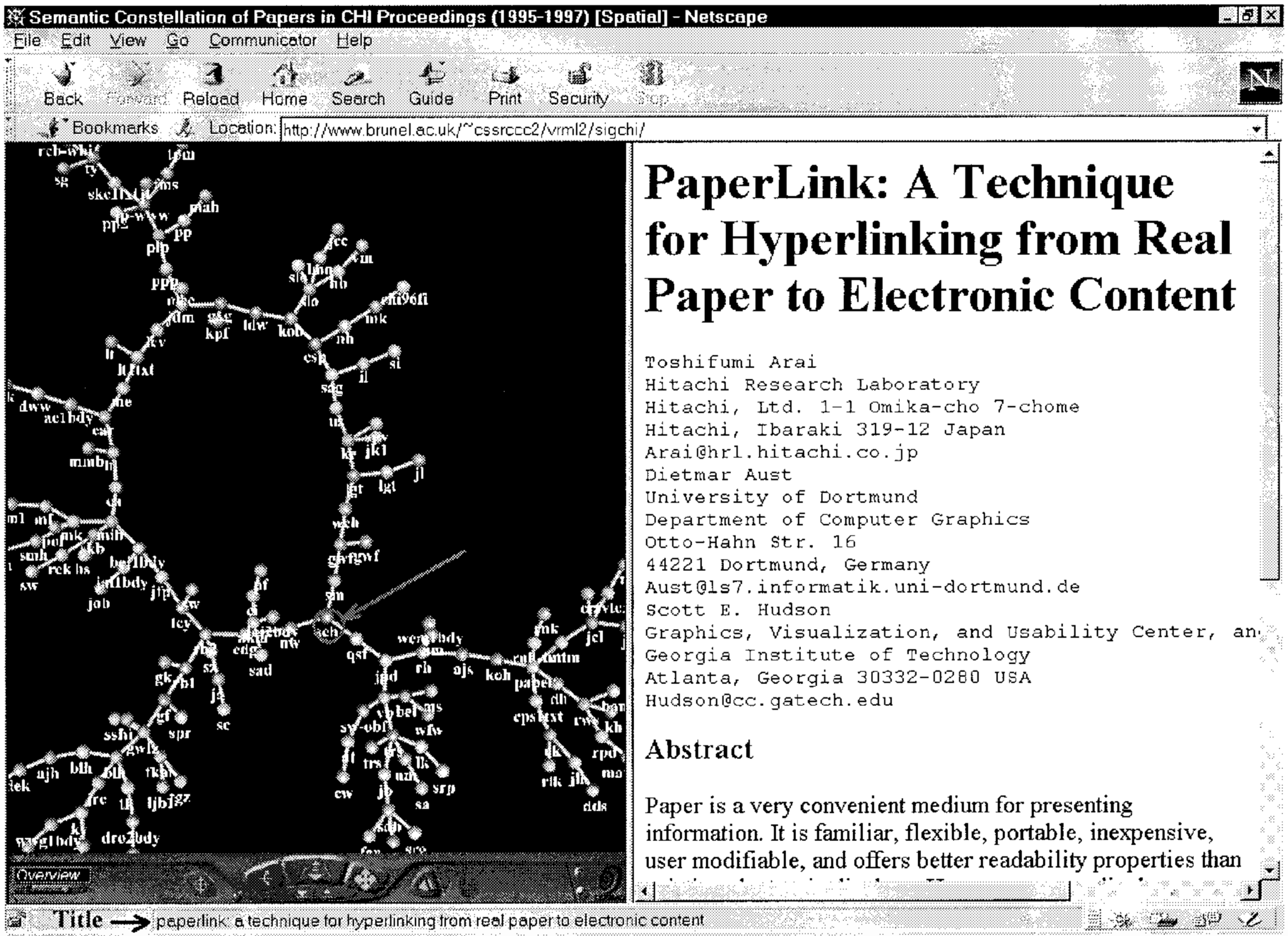}
    \caption{3D virtual information space used in Chen experiments.}
    \label{fig:fig12}
\end{figure}

Chen \cite{chen2000individual} combined spatial and semantic architectures to understand the individual differences in navigation of spatial-semantic virtual environments. The individual differences for a human-computer interaction (HCI), as defined in this work, encompass three cognitive factors: spatial ability (VZ-2), associative memory (MA-1), and visual memory (MV-1). These are reference cognitive tests developed by Educational Testing Service (ETS)\footnote{\url{https://www.ets.org}} as we have previously discussed. The virtual environment developed for the experiments presented documents from ACM CHI proceedings (1995-1997) in 3D semantic space by Latent Semantic Indexing (LSI) \cite{deerwester1990indexing} and is shown in Figure \ref{fig:fig12}. The work appeared to be motivated by information foraging theory \cite{pirolli1995information} which sought to formalize the notion of trade-offs between information gain and user load. Thus, by implementing a spatial-semantic metaphor to enter an unknown information space characterized by geometric and topological properties, similar documents clustered together within the space would theoretically reduce cognitive load. Moreover, the use of the term 'cognitive load' is subjective in that individual differences in verbal and spatial ability are crucial in understanding how to carefully map cognitive load to navigation strategies.  

Building upon the question of \textit{how people reach their destination using maps}, Chen asked the following question: \textit{What is the impact of an understanding of the latent semantics on individuals search strategies?} The first experiment aimed to explore the relationship between associative memory and visual memory, against visual navigation performance. There were ten subjects in the study with age ranges of 25-40, all with computer literacy. Associative memory and visual memory scores (MA-1, MV-1, respectively) were obtained 24 hours before the experiment. Subjects were allowed as much time as they wanted to learn the interface and controls. They were given two search topics and assigned the task of finding as many relevant papers within 15 minutes for the first topic. For the second topic, they were assigned to stop once they found 5 relevant papers; no time limit was given. At the end of the first search task, subjects were asked to draw the spatial layout of the retrieved information space. After the second search task, they were asked to name the cluster of articles in the information space; this is known as abstraction. 

\begin{figure}[htb]
    \centering
    \includegraphics[width=7cm]{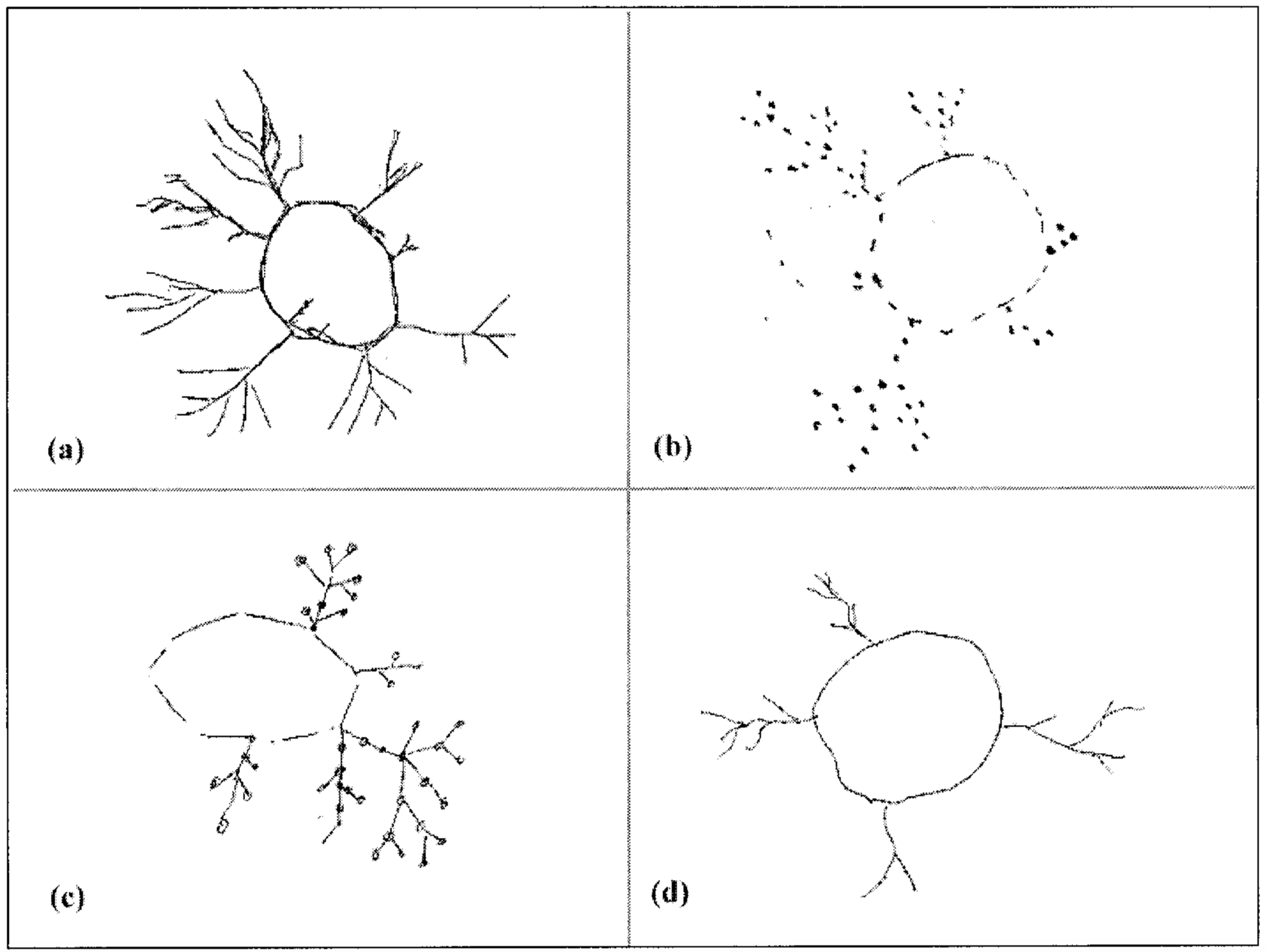}
    \caption{Participant abstractions of the information space in Chen experiment. (a-d) represents descending performance.}
    \label{fig:fig13}
\end{figure}

To create a baseline of performance, Chen extracted keywords from each task description to formulate a search query. The top 20 articles returned by LSI were used as the benchmark list of relevant articles. Thus, performance scores were measured by the proportion of top 20 articles a subject decided was relevant. For visual memory correlations with drawings, all subjects had widely varying levels of detail in their diagrams which provided an indication of how spatial memory may be influenced not only by the interface design of a virtual environment but also by individual differences in cognitive ability and navigation strategies. Additionally, abstractions of the space did not include coherent topical themes as some users leveraged metaphors, personal phrases, and other non-sensical words, as opposed to deliberately attempting to recall and characterize the document clusters they navigated. An example of the user abstractions is shown in Figure \ref{fig:fig13}. The results for the first experiment indicated that associative memory (MA-1) was most strongly correlated (p = 0.006) with recall scores, visual memory had no significant interactions with recall.

The second experiment aimed to explore the relationship between individual differences in spatial ability (VZ-2) and associative memory (MA-1) for two user interfaces (sequential and spatial). The twelve subjects in this study were either Ph.D. students, researchers, or academic staff in the authors department. Again, a pretest of spatial ability and associative memory were obtained. Subjects were asked to find as many articles related to four topics as they could in 10 minutes. For the first two topics, they used the spatial system. For the remaining two topics, they used the sequential system. After each topical search, subjects were given a questionnaire about their understanding of the topic. For relevance ratings, two schemes were used. In the first, pooled answers from all subjects was used to rate the proportional relevance of a document. The second rating scheme was based on query-document relevance ratings from LSI, exactly similar to the scheme in the first experiment. The relevance rating schemes by LSI and pooled answers were significantly correlated for both recall and precision, thus LSI-based evaluation became the benchmark for downstream analysis. 

Upon examination of cognitive abilities and search performance (recall), spatial ability and associative memory were strongly correlated for only one search topic while no other statistically significant interactions were observed. The main effects of cognitive ability were also examined by a multivariate procedure. The dependent or response variable was LSI-based recall and precision scores. On-line experience (years) was the most significant. However, no statistically significant effects were found for spatial ability and associative memory. This finding led to a univariate analysis of on-line experience. The author observed that the impact of on-line experience and search performance (recall) was statistically significant only for the sequential system.

\section{Discussion}

The studies that have been surveyed in this work, share the following general approach.
\begin{itemize}
  \item A custom multi-dimensional and spatial information retrieval system was experimented against a one-dimensional and sequential information retrieval system.
  \item Subjects were typically recruited from undergraduate/graduate programs from various academic disciplines.
  \item Psychometric evaluations were administered to assess verbal and spatial abilities.
  \item A learning phase for novel system architectures and various components prior to the experiment was allowed.
  \item Series of information-seeking tasks were clearly defined.
  \item Document corpora were not specifically tailored to domain expertise of subjects.
  \item Between-system retrieval performance was consistently defined by time, precision, and recall.
\end{itemize}
The studies differed in a number of ways as well.
\begin{itemize}
    \item On the issue of multiple dimensions, some system architectures were 2-dimensional, while others had both 2-dimensional and 3-dimensional implementations. 
    \item Among the recruited subjects, some had academic backgrounds in psychology, library science, information science, or computer science. All of which have elements of formal training in information retrieval architectures, potentially biasing the results.
    \item Only one study explicitly described their psychometric evaluations as \textit{controlled associations} or \textit{paper folding}, as opposed to \textit{verbal} or \textit{spatial} reasoning evaluation of which their are many kinds.
    \item Each study implemented varying lengths of learning phases prior to experimentation with novel system architectures, which potentially biases the results.
    \item Only two studies incorporated information processing evaluation (as an addition to retrieval performance) defined by abstracting information spaces or instructing subjects to describe what they retrieve(d).
\end{itemize}

Based on my observations it seems that cognitive and behavioral aspects of \textit{Sequential/Spatial} retrieval experiments demonstrate that some form of spatial encoding of information may reduce time and complexity of retrieval, while increasing information comprehension, in two types of instances – for individuals with higher than average spatial reasoning ability and/or after an undefined amount of time learning a new spatial system regardless of spatial reasoning ability. Moreover, while a user can certainly grasp the semantic relationships within a hierarchical or sequential information retrieval system over time, I am compelled to consider that human mental maps of semantic structure may be interpreted significantly earlier with spatially encoded supplementary information the longer that such a system is used.
 
Additionally, the evidence I have surveyed suggests that measures such as precision, recall (although differences were observed), and time, are perhaps not the most effective way of comparatively evaluating \textit{Sequential/Spatial} for information retrieval performance, as spatial information retrieval systems are more suited toward information foraging or sense-making, exploration, and learning behaviors than look-up or transactional procedures that rely heavily upon such measurements. It is important to understand information processing, in conjunction with or in complete replacement of, information retrieval performance. Thus, measurements of information space comprehension whether predicted by a machine or by a human assessor may lead to more insight on the human factors involved while processing information presented by spatially encoded systems.

It must not be overlooked that there will be inherent difficulty with interpreting spatial-semantics of spatial systems as they are non-traditional and users with particular cognitive abilities, may impose their mental model within an information space and arrive at information seeking conflict or tension. Moreover, direct manipulation issues can arise as subjects that are not comfortable operating a 2D input (computer mouse) to navigate a 3D virtual space will under-perform whereas the same spatial information placed in a lower-dimensional 2-D information space may in-fact result in higher performance and comprehension.

These studies all align very nicely with information foraging theory~\cite{pirolli1995information} which sought to formalize the notion of trade-offs between information gain and user load. Moreover, the use of the term ’cognitive load’ is subjective in that individual differences in verbal and spatial ability are crucial in understanding how to carefully map cognitive load to navigation strategies. Perhaps with the addition of physiological measures such as heart-rate, blood-pressure, galvanic skin conductance (perspiration) and neurophysiological measures such as eye-tracking, electroencephalography (EEG), and functional Magnetic Resonance Imaging (fMRI)~\cite{Gwizdka:2017:NCB:3020165.3022165}.

\section{Conclusion}

It is apparent that a gold standard evaluation for \textit{Sequential/Spatial} does not yet exist although each study contributed toward a full-spectrum understanding of sequential versus spatial information seeking/processing. While all experiments examined information seeking tasks, the experiments also had slightly different approaches. In order to understand the mechanics of spatial-semantic information retrieval, from the human perspective, information-seeking tasks and their choice of measurement need to be carefully constructed that more clearly isolate perceptual, cognitive, motor, and retrieval performance, components of information seeking. Moreover, supplementary data needs to be collected that demonstrates the relationship of the aforementioned components, along with mapping precise cognitive states and user behavior, to the effectiveness of spatial-semantic information retrieval tools.

\bibliographystyle{ACM-Reference-Format}
\bibliography{comp_references.bib}

\end{document}